# New approaches for increasing the reliability of the *h* index research performance measurement

Lutz Bornmann,[a,1] Rüdiger Mutz,[a] and Hans-Dieter Daniel[a,b]

[a] Professorship for Social Psychology and Research on Higher Education, ETH Zurich, Zähringerstr. 24, 8092 Zurich, Switzerland
[b] Evaluation Office, University of Zurich, Mühlegasse 21, 8001 Zurich, Switzerland

[1]Corresponding author:
Dr. Lutz Bornmann
Professorship for Social Psychology and Research on Higher Education
ETH Zurich
Zähringerstr. 24
8092 Zurich
Switzerland
Tel.: +41 (0)44 632 48 25
Fax: +41 (0)44 632 12 83
E-mail: bornmann@gess.ethz.ch

# Abstract

In the year 2005 Jorge Hirsch introduced the *h* index for quantifying the research output of scientists. Today, the *h* index is a widely accepted indicator of research performance. The *h* index has been criticized for its insufficient reliability – the ability to discriminate reliably between meaningful amounts of research performance. Taking as an example an extensive data set with bibliometric data on scientists working in the field of molecular biology, we compute $h^2$ lower, $h^2$ upper, and sRM values and present them as complementary approaches that improve the reliability of the *h* index research performance measurement.

# 1    Introduction

Physicist Jorge Hirsch (1) introduced an indicator for quantifying the research output of scientists that has ever since been discussed and studied theoretically and empirically in a number of disciplines (2-3). Hirsch's *h* index was proposed as a better alternative to other bibliometric indicators (such as number of publications, average number of citations, and sum of all citations) (4). It is based on a scientist's lifetime citedness (5), which incorporates productivity as well as citation impact: "A scientist has index *h* if *h* of his or her $N_p$ papers have at least *h* citations each and the other ($N_p − h$) papers have ≤*h* citations each" (1, p. 16569). All works by a scientist having at least *h* citations are called the 'Hirsch core' (6); these are the publications within a scientist's publication list that have the greatest visibility (or greatest impact) (7).

Today, the *h* index is a widely accepted indicator of research performance; the *h* index is computed automatically in the Web of Science (provided by Thomson Reuters, Philadelphia, PA) citation reports on a publication list (8). A number of studies showed that a scientist's *h* index correspond to peer judgments (2-3) and thus has convergent validity. The most-often expressed criticism of the *h* index (and which led to the development of numerous variants of the *h* index) is applicable not only to the *h* index itself but also to bibliometric indicators generally (the problems of field dependency, self-citations, and multi-authorship). Criticism specific to the *h* index is much more rarely to be found in the literature. But as it concerns its insufficient reliability (9-10), or the ability to discriminate reliably between meaningful amounts of research performance, it is fundamental and noteworthy: For one, it is said that a single *h* index value does not yield an reliable picture of the research performance of a scientist; additional data are necessary. For another, a scientist's *h* index value is said to differ more of less from the 'true' value of the number of his/her core publications (that is, those publications with the greatest visibility, see above).

## 2   Criticism of the reliability of the *h* index

The *h* index value results from the distribution of citations (11) over a scientist's rank-ordered publications. This distribution contains the complete information on the productivity and the citation impact of a scientist. However, the *h* index captures only "a small amount of information about the distribution of a scientist's citations" (11, p. 2) and discards "almost all the detail of citation records" (11, p. 14), for which reason it is "not applicable to the general body of researchers" (12, p. 16). Scientists that have similar *h* index values can have very different research performance types "described in terms of the production of scientific papers and their quality (as assessed by citations)" (13, p. 382): "Think of two scientists, each with 10 papers with 10 citations, but one with an additional 90 papers with 9 citations each; or suppose one has exactly 10 papers of 10 citations and the other exactly 10 papers of 100 each. Would anyone think them equivalent?" (11, p. 13). In section 4 below, we will present $h^2$ lower and $h^2$ upper, which shed light on the amount of information about the distribution of a scientist's citations not captured by the *h* index.

The problem with the way in which the *h* index combines publication and citation numbers has been described as follows: "The problem is that Hirsch assumes an equality between incommensurable quantities. An author's papers are listed in order of decreasing citations with paper *i* having $C(i)$ citations. Hirsch's index is determined by the equality, $h = C(h)$, which posits an equality between two quantities with no evident logical connection" (14, p. 377). The equality $h = C(h)$ is viewed as an oversimplification (15) and as arbitrary: "Hirsch could equally well have defined the *h*-index as follows: A scientist has *h*-index *h* if *h* of his *n* papers have at least 2h citations each and the other n − h papers have fewer than 2(h + 1) citations each. Or he could have used the following definition: A scientist has *h*-index *h* if *h* of his *n* papers have at least h/2 citations each and the other n − h papers have fewer than (h + 1)/2 citations each. A priori, there is no good reason why the original definition of the *h*-index would be better than these two alternative definitions and other similar definitions. Hence, the *h*-index can be seen as a special case of a more general research performance measure. The *h*-index is obtained from this more general measure by setting a parameter to an arbitrarily chosen value" (8, pp. 263-264). In section 5, we will present a simple approach whereby this parameter is not set arbitrarily but is instead estimated based on the distribution of a scientist's publication and citation data. This estimated value, which we call the sRM value, sheds light on the 'true' value of the number of a scientist's publications with the greatest visibility (that is, his/her 'true core'), and the difference between the *h* index and this estimated, 'overarching' research performance measure can be determined.

## 3    Description of the data set

We investigate the *h* index and present approaches that increase the reliability of the *h* index research performance measurement using a data set containing the publications of applicants to the Young Investigator Programme (16-18). This program of the European Molecular Biology Organization (EMBO) in Heidelberg, Germany, has been supporting outstanding young group leaders in the life sciences in Europe since 2000 (see http://www.embo.org/yip/index.html; accessed: July 10, 2009). The program targets researchers who have been leading their first independent laboratory in a European Molecular Biology Conference (EMBC) Member State (see http://embc.embo.org/, accessed: July 10, 2009) normally not more than four years before applying to the program. The study examined publication and citation data for 297 applicants to the EMBO Young Investigator Programme from the years 2001 and 2002. These applicants published a total of 6,087 papers (articles, letters, notes, and reviews) prior to submitting their applications (publication window: from 1984 to the application year in 2001 or 2002). These papers received an average of 46.56 citations (median=23) (citation window: from publication year to the beginning of 2007). The applicants' *h* index values were on average 13.13 (arithmetic average, median=13) and range from 1 (minimum) to 34 (maximum).

# 4 Calculation of *h²* lower and *h²* upper

Fig. 1 shows for three applicants the distribution of citations over each applicant's publication set. The area of a distribution tallies with the applicant's total citation counts. As a rule, the citation distribution for a larger number of publications is right-skewed, distributed according to a power law (19). In a publication set there are mostly a few highly cited papers and many hardly cited papers. As the distribution for scientist A shows, the *h* index captures only a small part of the publication and citation data, if the distribution is right-skewed. The *h* index refers to the area *h\*h* and does not take into consideration the areas starting at *h* citations (we will call this *h²* upper) or starting at *h* papers (we will call this *h²* lower). For this reason, different scientists for whom the citation frequencies are distributed very differently right-skewed to their publications can have the same *h* index (see scientists A, B, and C in Fig. 1). The area proportions *h²* lower, *h²*, and *h²* upper are defined as follows:

$$h^2 \text{ upper} = \frac{\sum_{j=1}^{h}(\text{cit}_j - h)}{\sum_{j=1}^{n}\text{cit}_j} \cdot 100 \qquad (1)$$

$$h^2 = \frac{h \cdot h}{\sum_{j=1}^{n}\text{cit}_j} \cdot 100 \qquad (2)$$

$$h^2 \text{ lower} = \frac{\sum_{j=h+1}^{n} \text{cit}_j}{\sum_{j=1}^{n} \text{cit}_j} \cdot 100 \qquad (3)$$

As the equations show, the research performance of two or more scientists could be compared by examining $h^2$ lower, $h^2$, and $h^2$ upper in percent of total citation counts. To determine $h^2$ upper in Web of Science, publications in a scientist's publication list (sorted by 'times cited') that have citation counts greater than the scientist's $h$ index value must be marked and for these publications a citation report produced to obtain the sum of citations for these publications (see 'sum of the times cited' in the report). If $h*h$ is subtracted from this sum, the result – given in percent of the total citation counts – is $h^2$ upper. To obtain $h^2$ lower, $h^2$ upper and $h^2$ must be subtracted from 100.

As Fig. 1 shows, the three applicants A, B, and C, who have the same $h$ index values (here 14), have very different values for $h^2$ lower, $h^2$, and $h^2$ upper. This indicates very different research performance types. Whereas $h^2$ lower for applicant A makes up about 3% of the entire area of the distribution, for applicant C this is 57%. We find the opposite for $h^2$ upper, which makes up 82% of the entire area of the distribution for applicant A but only 10% for applicant C. Applicant A is a scientist who has rather few but very highly cited publications. Cole and Cole (13) call this type of scientist *perfectionists*, who publish "comparatively little but what they do publish has a considerable impact on the field" (p. 382). Applicant B can be called a *prolific scientist* following Cole and Cole (13), "in the dual sense of producing an abundance of papers which tend also be fruitful" (p. 382), and Applicant C is a *mass producer*, who publishes "a relatively large number of papers of little consequence" (p. 382).

Table 1 shows the area proportions $h^2$ lower, $h^2$, and $h^2$ upper for applicants that have similar $h$ index values. For example, for applicants that have an $h$ index value of 10 or 11 ($n=40$), $h^2$ covers on average 20% of the area of the distribution of citations. The minimum value for the applicants is on average 6% and the maximum value 44%. This means, for one, that the variability in the covering of the citation distribution by $h^2$ is very high, and it shows, for another, that in this $h$ index subgroup there is no applicant for whom $h^2$ makes up at least one-half of the entire area. As the percentages in the column Total show, $h^2$ lower generally makes up 7% of the area. Across all $h$ index subgroups, $h^2$ refers to only about one-fourth of the entire area. The by far greatest share of the area (70%) generally traces back to $h^2$ upper. Hence, especially the area of highly cited papers in a distribution will be depicted only insufficiently by $h^2$ (and with it, by the $h$ index value).

Altogether, on the basis of the three graphs in Fig. 1 and the minimum and maximum values in Table 1 it is clearly visible that for applicants *having the same h index value, $h^2$* covers a very different proportion of the area of the individual citation distribution. Among the applicants having similar $h$ index values, there are therefore very different research performance types, which can be identified through the additional values $h^2$ lower and $h^2$ upper, however.

## 5    Calculation of sRM values

According to Seglen (5) "each individual scientist may constitute a 'microfield' with a characteristic citation probability determined by that individual's research profile" (p. 637). Accordingly, the 'true' value of the number of publications in the core (that is, the scientist's most visible publications) can result only from the citation distribution over his/her publications. As we showed in section 4, however, with very different citation distributions for scientists the $h$ index values can still be very similar. Thus, the $h$ index only insufficiently captures the complete distribution in the calculation. But with the use of the segmented regression model (sRM) there is a way to determine the 'true' value (we will call it the sRM value) of the number of publications in the core based on the individual citation distribution.

Typically, citation counts decrease and cumulated citation counts increase from a high rank publication to a low rank publication, with a steep slope in the first part ('core' publications with high visibility) and a flat slope in a second part (publications with low visibility) (see Fig. 2). To statistically model the citation transition zone between the first and the second part there is a need for two separate regressions to obtain a reasonable fit for the whole citation distribution (20-21). The first part can be best described by a quadratic curve, the second part by a linear curve. With segmented regression a statistical model is given that is able to simultaneously estimate the parameters of the two curves and the joint point between publications with high impact (the 'true core') and publications with low impact (22-23).

The following sRM for $y_j$ was assumed, whereby $z_0$ is the break point between publications in the first and publications in the second part of the distribution:

if $x_j < z_0$

$$y_j = b_0 + b_1 x_j + b_2 x_j^2 + e_j \qquad e_j \sim N(0, \sigma_e^2)$$

otherwise

$$y_j = b_0 + b_1 z_0 + b_2 z_0^2 + b_3 (x_j - z_0) + e_j \qquad e_j \sim N(0, \sigma_e^2) \qquad (4)$$

The x values (ranked publications) for j range from 1 to k. The $z_0$ value is defined as the maximum of the quadratic function:

$$z_0 = \frac{-b_1}{(2b_2)} \qquad (5)$$

The sRM can be estimated using non-linear least squares. The statistical software SAS, for instance, offers a procedure called NLIN, which allows computation of the parameters with Gauss-Newton iteration (24) (see the SAS program in Appendix A). The size of the residual variance ($\sigma^2_e$), or the proportion of explained variance to total variance (R²), gives some evidence for the amount of model fit. This sRM is applicable to a scientist's publication set, if (i) two different parts in the citation distribution can be clearly distinguished (this is mostly the case for scientists, as their distributions of publications' citedness are found to be very skewed, 5), (ii) the algorithm converges, (iii) R² is high (>.90), (iv) the breakpoint lies within the range of publications, and (v) there are in sum at least 15 to 20 publications for one scientist (25). These requirements should not be seen as disadvantages of the model in application; the *h* index, too, should not be computed for every scientist (for instance, when comparing scientists having low publication and citation counts, it is not very meaningful).

For subgroups of applicants having similar *h* index values Table 2 shows the arithmetic means, standard deviations, and minimum and maximum values of sRM values. As the means in the table show, when there is an increase in the *h* index values there is also an increase in the sRM values. At the aggregate level, sRM and *h* correspond. However, the sRM values are very spread out around the means in Table 2. Across all applicants the standard deviation of sRM values is 8.45. Thus, among the applicants with an *h* index of 8 or 9, there are scientists with sRM values ranging from 2.95 to 27.59. The other *h* index subgroups show a similar picture. This means that scientists who based on the *h* index value should have a similar research performance have a very differently sized core of most visible publications – when the individual citation distribution is taken as a basis.

The *h* index value of a scientist depends on the field in which he/she publishes. This dependency is mainly due to differences in the expected citation rates in the fields: "There will be differences in typical *h* values in different fields, determined in part by the average number of references in a paper in the field, the average number of papers produced by each scientist in the field, and the size (number of scientists) of the field … Scientists working in non-mainstream areas will not achieve the same very high *h* values as the top echelon of those working in highly topical areas" (1, p. 16571). Because the sRM value results from the individual citation distribution of a scientist and does not refer to the absolute number of citations for the individual publications, it is not dependent on the expected citation rate in a field. The sRM values of scientists in different fields can therefore be compared without normalization – if in those fields the average productivity of the scientists (in journal papers) is similar.

This advantage of the sRM value over the *h* index is also a drawback, however, in that the sRM value says nothing about the number of citations of the most visible publications. Whereas the *h* index value provides an indication of the absolute number of citations of the publications in the 'Hirsch core' (at least *h* citations each), this information is lacking in the sRM value.

## 6    Discussion

In addition to the advantages of the *h* index (such as simple calculation), a number of disadvantages have been named in recent years (2, 26), which has led to the development of numerous *h* index variants (27). As we showed in several publications (27-29), only some of these variants are associated with an incremental contribution for evaluation purposes – the *h* index and many of the variants are highly correlated. We therefore do not think it would be wise to develop further variants of the *h* index in future, but it is useful to complement the *h* index with additional information in order to obtain a more complete and more reliable picture of the research performance of a single scientist. In this study we presented $h^2$ lower, $h^2$ upper, and the sRM value as approaches that are easily computed and that provide high information content. When evaluating a single scientist we recommend, in addition to the *h* index, a look at the citation distribution over the scientist's publications. When evaluating the research performance of a larger number of scientists, where this cannot be done for each individual case, the approaches presented here could be used. But as a fundamental principle, academic 'age' of and the field in which the different scientists work should always be taken into account (3).

Using $h^2$ lower, $h^2$ upper, and the sRM values, we can "best derive useful information from citation data" (11, p. 2), as recommended by the International Mathematical Union (Berlin, Germany). These approaches make possible insight into "the detail of citation records" (11, p. 14), which will bring us a step closer to solving "the problem of comparing or combining different h-indexes" (30, p. 369).

# Acknowledgements

The authors would like to thank Dr. Gerlind Wallon, deputy director of the European Molecular Biology Organization (EMBO), and Dr. Anna Ledin, working as a scientific secretary for the Royal Swedish Academy of Sciences in Stockholm (and former at EMBO), for providing the bibliographic data on the applicants to the EMBO Young Investigator Programme.

Fig. 1. Citation distributions of three applicants (A, B, and C) with the same $h$ index

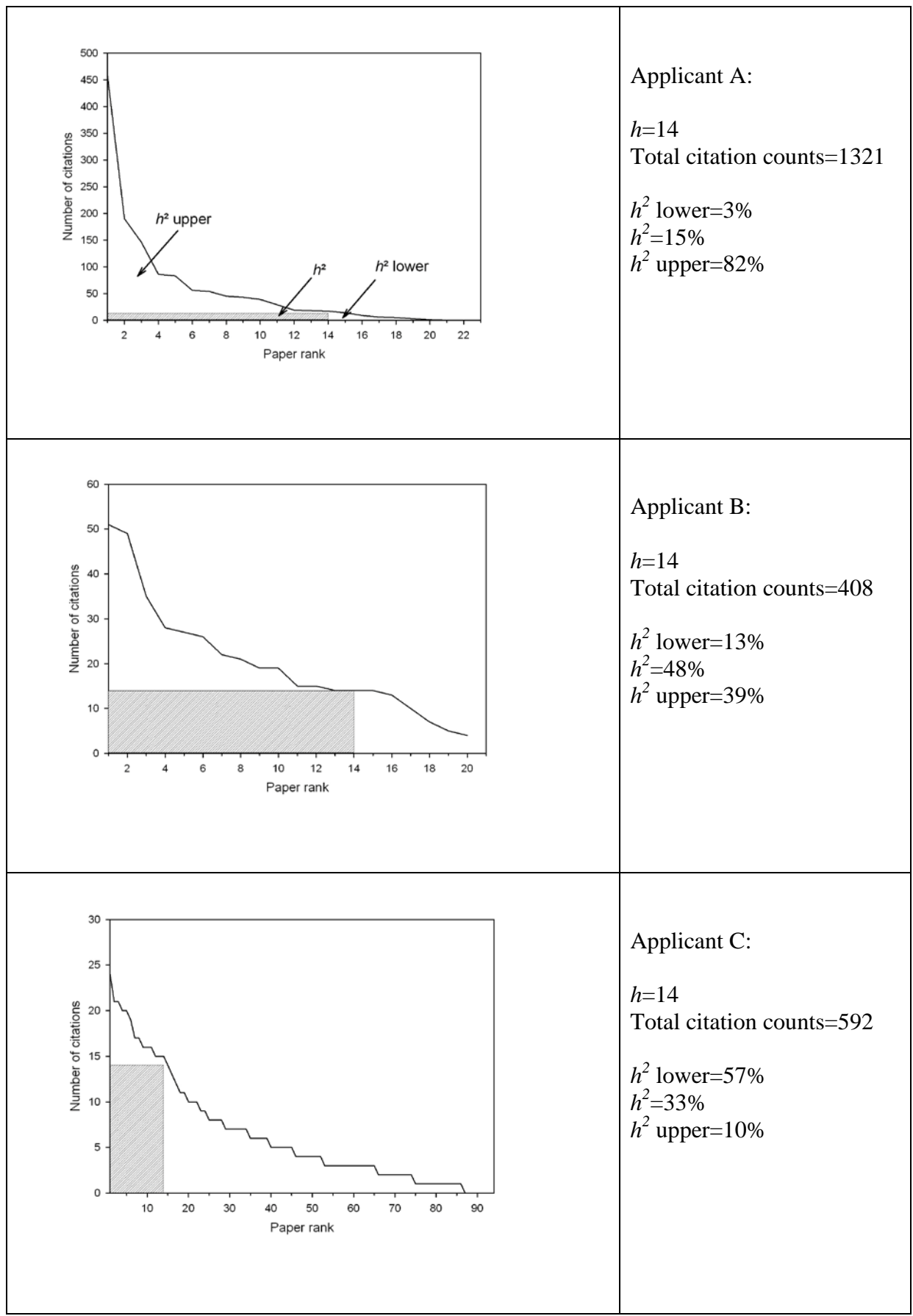

Applicant A:

$h$=14
Total citation counts=1321

$h^2$ lower=3%
$h^2$=15%
$h^2$ upper=82%

Applicant B:

$h$=14
Total citation counts=408

$h^2$ lower=13%
$h^2$=48%
$h^2$ upper=39%

Applicant C:

$h$=14
Total citation counts=592

$h^2$ lower=57%
$h^2$=33%
$h^2$ upper=10%

Table 1.
Arithmetic mean, standard deviation, minimum and maximum of $h^2$ lower, $h^2$ and $h^2$ upper by different $h$ index values of the applicants (percentages)

| $h$ index value | Number of applicants | Arithmetic mean | Standard deviation | Minimum | Maximum |
|---|---|---|---|---|---|
| **$h^2$ lower** | | | | | |
| <=7 | 32 | 4 | 5 | 0 | 18 |
| 8 - 9 | 47 | 6 | 10 | 0 | 50 |
| 10 - 11 | 40 | 5 | 6 | 0 | 27 |
| 12 - 13 | 50 | 6 | 7 | 0 | 39 |
| 14 - 15 | 43 | 8 | 11 | 0 | 57 |
| 16 - 17 | 30 | 9 | 9 | 0 | 43 |
| 18 - 19 | 22 | 6 | 6 | 0 | 21 |
| >=20 | 33 | 9 | 7 | 1 | 25 |
| Total | 297 | 7 | 8 | 0 | 57 |
| **$h^2$** | | | | | |
| <=7 | 32 | 17 | 14 | 4 | 64 |
| 8 - 9 | 47 | 22 | 12 | 3 | 60 |
| 10 - 11 | 40 | 20 | 8 | 6 | 44 |
| 12 - 13 | 50 | 25 | 10 | 11 | 47 |
| 14 - 15 | 43 | 25 | 9 | 11 | 48 |
| 16 - 17 | 30 | 25 | 8 | 9 | 45 |
| 18 - 19 | 22 | 24 | 8 | 10 | 39 |
| >=20 | 33 | 26 | 9 | 8 | 45 |
| Total | 297 | 23 | 10 | 3 | 64 |
| **$h^2$ upper** | | | | | |
| <=7 | 32 | 79 | 17 | 29 | 96 |
| 8 - 9 | 47 | 72 | 19 | 10 | 97 |
| 10 - 11 | 40 | 76 | 13 | 29 | 94 |
| 12 - 13 | 50 | 69 | 15 | 34 | 88 |
| 14 - 15 | 43 | 66 | 17 | 9 | 88 |
| 16 - 17 | 30 | 67 | 15 | 32 | 91 |
| 18 - 19 | 22 | 70 | 12 | 42 | 90 |
| >=20 | 33 | 64 | 14 | 41 | 91 |
| Total | 297 | 70 | 16 | 9 | 97 |

Fig. 2. sRM value and *h* index value for one applicant with *h*=14

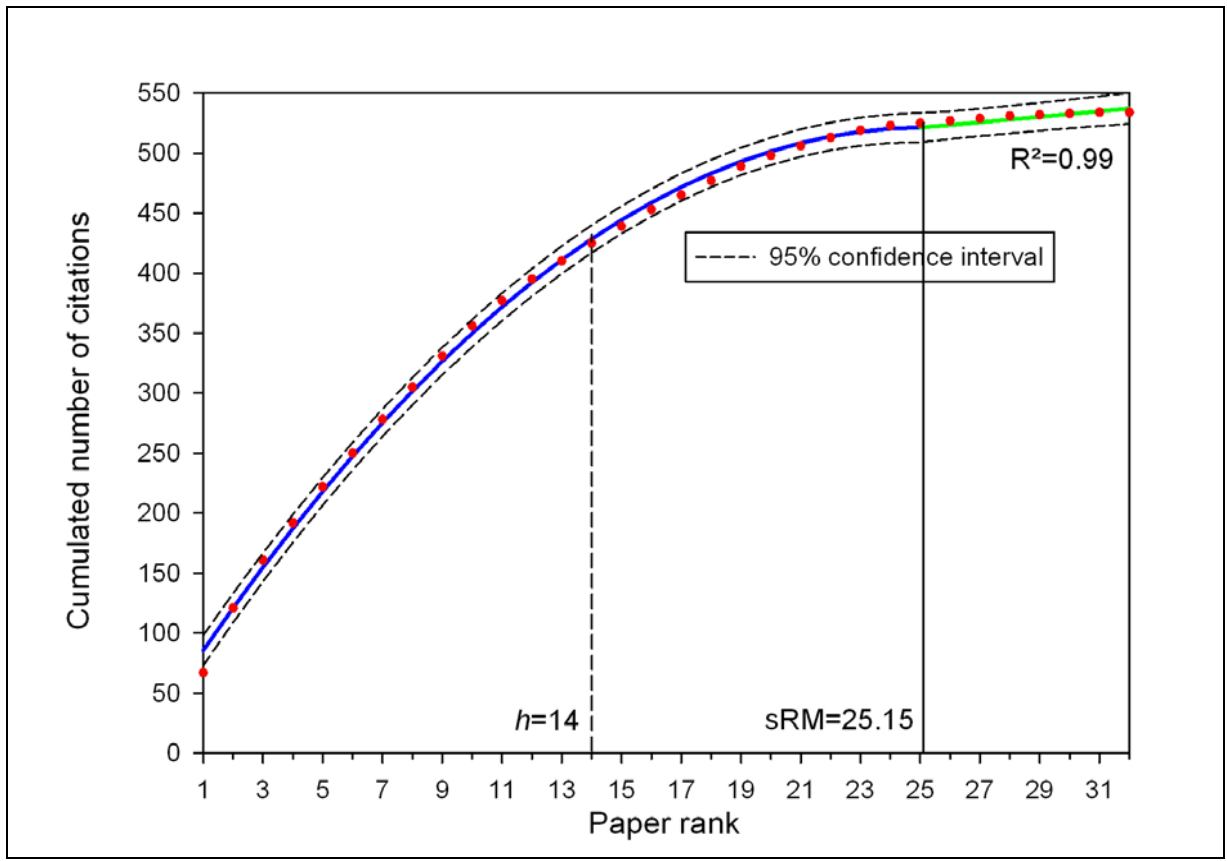

*Note*. The red dots are the applicant's cumulated number of citations. The blue line is the fitted quadratic curve (with 95% confidence interval), and the green line is the fitted linear curve (with 95% confidence interval). The $R^2$ of 99% indicates that the fitted values explain the applicant's cumulated number of citations almost completely. The *h* index value (14), which is clearly lower than the sRM value (25.15), shows that the number of highly cited papers of the applicant is underestimated by the *h* index by about 10 publications.

Table 2.
Arithmetic mean, standard deviation, minimum and maximum of sRM values by different *h* index values of the applicants

| *h* index value | Number of applicants | Arithmetic mean | Standard deviation | Minimum | Maximum |
|---|---|---|---|---|---|
| <=7 | 20 | 5.94 | 1.87 | 2.83 | 9.41 |
| 8 - 9 | 37 | 10.19 | 4.83 | 2.95 | 27.59 |
| 10 - 11 | 35 | 9.93 | 2.69 | 4.41 | 14.72 |
| 12 - 13 | 35 | 13.86 | 6.29 | 3.53 | 41.90 |
| 14 - 15 | 36 | 15.95 | 8.75 | 7.95 | 60.73 |
| 16 - 17 | 28 | 18.73 | 6.87 | 9.94 | 43.77 |
| 18 - 19 | 19 | 18.79 | 5.24 | 9.05 | 31.67 |
| >=20 | 31 | 24.83 | 10.13 | 8.20 | 57.39 |
| Total | 241 | 14.75 | 8.45 | 2.83 | 60.73 |

*Note*.
sRM values could not be computed for 56 of the total 297 applicants, because one or more requirements for calculation of the sRM were not met.

Appendix A

SAS program for calculating the sRM:

```
proc nlin data=dataset;
parms a = 35 to 70 by 25
      b = 20 to 80 by 10
      c = 4 to -4 by -0.50
      d=  2 to -2 by -0.25;
x0=-0.5 * b/c;
model cumcitation=ifn(art<x0, a+b*art+c*art*art,a+b*x0+c*x0*x0+d*(art-x0));
run;
```